\documentstyle[12pt,a4,epsfig]{article}

\begin{document} 
\bibliographystyle{unsrt} 
\baselineskip= 16pt 
 \newcommand \be {\begin{equation}}
\newcommand \ee {\end{equation}}
 \newcommand \ba {\begin{eqnarray}}
\newcommand \ea {\end{eqnarray}}

\begin{titlepage}
\title{Aging in glassy systems: new experiments, simple models, and open questions}
\author{J.P. Bouchaud\\ Service de
Physique de l'Etat 
Condens\'e, CEA Saclay,\\ 91191 Gif sur Yvette Cedex, France \
}
\maketitle
\begin{abstract}
We review the most striking experimental results
on aging in a variety of disordered systems, which reveal similar features but
also important differences. We argue that a generic model that reproduce
many of these features is that of {\it pinned defects} in a disordered environment, hopping between different metastable states. 
The fact that energy barriers grow with the size of the reconforming regions
immediately leads to a strong hierarchy of time scales. In particular, long wavelength glassy, aging modes and short wavelength equilibrated modes coexist in the
system and offer a simple mechanism to explain the rejuvenation/memory effect,
which does not rely on `chaos' with temperature. 
These properties can be discussed within simplified models (such as the Sinai model) where the dynamics of the whole pinned object is reduced to that of its center of mass in a disordered potential. 
While the experiments in random ferromagnet-like systems can be satisfactorily accounted for, the correct picture for the aging dynamics of spin-glasses is still missing. Following recent ideas of Houdayer and Martin, we
propose the idea that the ordered phase a spin-glass contains a large number of pinned, zero tension walls. Finally, we briefly discuss the status of the mean-field theories
of aging.
\end{abstract}

\end{titlepage}

\maketitle
\section{Introduction}

A great variety of systems exhibit slow `glassy' dynamics \cite{Review}: glasses of all kinds,
but also spin-glasses or dipolar glasses \cite{struik,Sitges}. Another very interesting class of systems is that
of pinned `defects' such as Bloch walls, vortices in superconductors,
charge density waves, dislocations, etc. interacting with randomly placed impurities
\cite{LeDou}. Another type of systems where surprisingly slow dynamics can occur are soft glassy materials, such as foams, dense emulsions or granular materials, which attracted much attention recently \cite{Sollich}.
Of particular interest is the so called {\it aging} phenomenon observed in the response function of these glassy systems. This response is either to a magnetic field
in the case of disordered magnets, to an electric field in the case of dipolar glasses, or to an applied stress in the case of, e.g. glassy polymers or dense emulsions. The basic phenomenon is that the
response is {\it waiting time} dependent, i.e. depends on the time $t_w$ one has waited in the
low temperature phase before applying the perturbation. Qualitatively speaking, these systems
{\it stiffen} with age: the longer one waits, the smaller the response to an external drive, as
if the system settled in deeper and deeper energy valleys as time elapses. More precisely, 
if a system is cooled in zero field and left at $T_1$ (below the glass temperature $T_g$ \footnote{Here the glass transition temperature is understood as the temperature below which the relaxation time becomes larger than the experimental time scales. This temperature may or may not correspond to a true phase transition.}), during a time $t_w$ before
applying an external field (or stress), then the time dependent magnetisation saturates 
after a time comparable to $t_w$ itself. In other words, the time dependent magnetisation (or strain)
takes the following form:
\be\label{Mt}
M(t_w+t,t_w) \simeq M_\infty(t) + M_{\sc ag}(\frac{t}{t_w}),
\ee
where $M_\infty(t)$ is a `fast' part,\footnote{Actually, $M_\infty(t)$ should be written 
$M_\infty(t/\tau_0)$, where $\tau_0$ is a microscopic time scale.} and $M_{\sc ag}$ is the slow, aging part. Analoguously, if a polymer glass is left at $T_1$ during a time $t_w$ before applying an external stress, the subsequent strains develop on a time scale given by $t_w$ itself. Note that the decomposition of the dynamics into a fast
part and a slow part often also holds {\it above} (but close to) $T_g$, where one speaks of $\beta-$relaxation and $\alpha-$relaxation, respectively. The time scale $\tau(T)$ for the $\alpha-$relaxation is however finite and waiting time independent above $T_g$. This time becomes effectively infinite for $T<T_g$, and is therefore replaced by the age of the system $t_w$. In other words, Eq. (\ref{Mt}) is 
the continuation in the glass phase of the precursor two-step relaxation commonly observed
above $T_g$ \cite{MCT}.

\subsection*{Aging in correlation functions}

One can also, at least numerically, observe aging in the {\it correlation functions}. For example, the dynamic structure factor of -- say -- a Lennard-Jones system, defined as:
\be
S_q(t_w+t,t_w) = \langle e^{i \vec q \cdot [\vec r(t+t_w)-\vec r(t_w)]} \rangle,
\ee
where $\vec r(t)$ is the position of a tagged particle at time $t$, exhibits 
interesting aging properties \cite{Barrat}. Note that the corresponding experiments are much harder to realize, since the measurement of the structure factor typically takes
several minutes or so. In order not to mix together different waiting times, one
should redo the experiments several times, heating back the system above $T_g$,
and cooling down again, until the number of runs is sufficient to obtain a 
good averaging for a {\it fixed} $t_w$. This point will be further discussed
below -- see Eq. (6).

In equilibrium, the correlation and the response function are related through the
{\it fluctuation-dissipation theorem}. Interestingly, this theorem does not hold
in general for the aging part of the correlation and response. A generalisation of this theorem has been provided in \cite{CuK,CuKP} in the context of some mean-field spin-glass models, where the true temperature of the system is replaced by an effective temperature, higher than that of the thermal bath (and possibly time dependent). Quite a lot of efforts have been devoted to measure this effective temperature
in glassy systems, either numerically \cite{FDT}, or experimentally \cite{Israeloff}.

\section{Different types of aging}

\subsection{Aging in the a.c. susceptibility}

Aging can also be seen on a.c. susceptibility measurements (response to an oscillating field at frequency $\omega$). These have the advantage that the
perturbing field can then be extremely small. On the other hand, since one
must wait for at least one period before taking a measurement, the time 
sector available in these experiments is confined to $\omega t_w > 1$ (corresponding, in the language of the time dependent magnetisation discussed
above, to the short time region $t < t_w$). The a.c. susceptibility typically takes the following form:
\be
\chi''(\omega,t_w) = \chi_{\infty}''(\omega) +  f(t_w) \chi_{\sc ag}(\omega)
\ee
where $\chi_{\infty}(\omega)$ is the stationary contribution, obtained after
infinite waiting time. Depending on the system, the aging part behaves
quite differently. For example, in spin glasses ({\sc sg}), both the functions $f$ and $\chi_{\sc ag}$ behave similarly, as a {\it power-law} \cite{Sitges}:
\be\label{sgaging}
\left. f(t_w) \chi_{\sc ag}(\omega)\right|_{\sc sg} \simeq A (\omega t_w)^{x-1}
\qquad x \sim 0.7 - 0.9.
\ee
This behaviour is the counterpart, in frequency space, of the $t/t_w$ scaling
reported above. Some dipolar glasses ({\sc dg}), close to a ferroelectric 
transition \cite{Levelut2,Levelut}, reveal a very different behaviour, since in this case one has:
\be
\left. f(t_w) \chi_{\sc ag}(\omega)\right|_{\sc dg} \simeq A t_w^{x-1}
\quad {\mbox {or}} \quad -B \log t_w,
\ee
i.e., an aging part which is nearly frequency independent. Glycerol, on the other
hand, shows an intermediate behaviour \cite{Nagel}: the aging part is frequency dependent, but the frequency dependence is weaker than the waiting time dependence. 

\subsection{Role of the thermal history}

If a system ages, then by definition it is out-of-equilibrium. Therefore, one
might worry that the properties that one measures at $T_1$ actually strongly
depends on the whole thermal history of the system. Here again, different 
systems behave very differently. A naive argument, based on the idea that 
thermal activation over high energy barriers plays a central role in aging, 
would suggest that the age of a system cooled very slowly before reaching $T_1$ 
should be much larger than the age of the same system, but cooled more rapidly.
In other words, cooling the system more slowly allows energy barriers to be surmounted more efficiently at higher temperatures, and thus brings the system
closer to equilibrium at $T_1$. This is precisely what happens for {\sc dg} 
systems \cite{Levelut2}. One can actually use non-uniform cooling protocols, where the
cooling rate is either slow or fast only in the vicinity of $T_g$, and then fixed to a constant value for the last few Kelvins -- see Fig. 1. In the case of {\sc dg}, one
sees very clearly that the cooling rate when crossing $T_g$ is the crucial 
quantity which determines how well the system is able to equilibrate \cite{Levelut2}. When
the system is cold, then the dynamics is essentially frozen on experimental time scales, and therefore the precise value of the cooling rate there is of minor
importance. 

Surprisingly, the situation is completely reversed for {\sc sg}. In these 
systems, the value of the cooling rate in the vicinity of $T_g$ is completely
irrelevant, and the observed a.c. susceptibility is to a large degree independent of the cooling rate, except for the very last Kelvins \cite{memchaos}. This is
illustrated in Fig. \ 1. The naive picture of a system crossing higher and
higher barriers to reach an optimal configuration therefore certainly needs 
to be reconsidered for these systems.

\vskip 0.2cm
\begin{figure}
\centerline{\hbox{\epsfig{figure=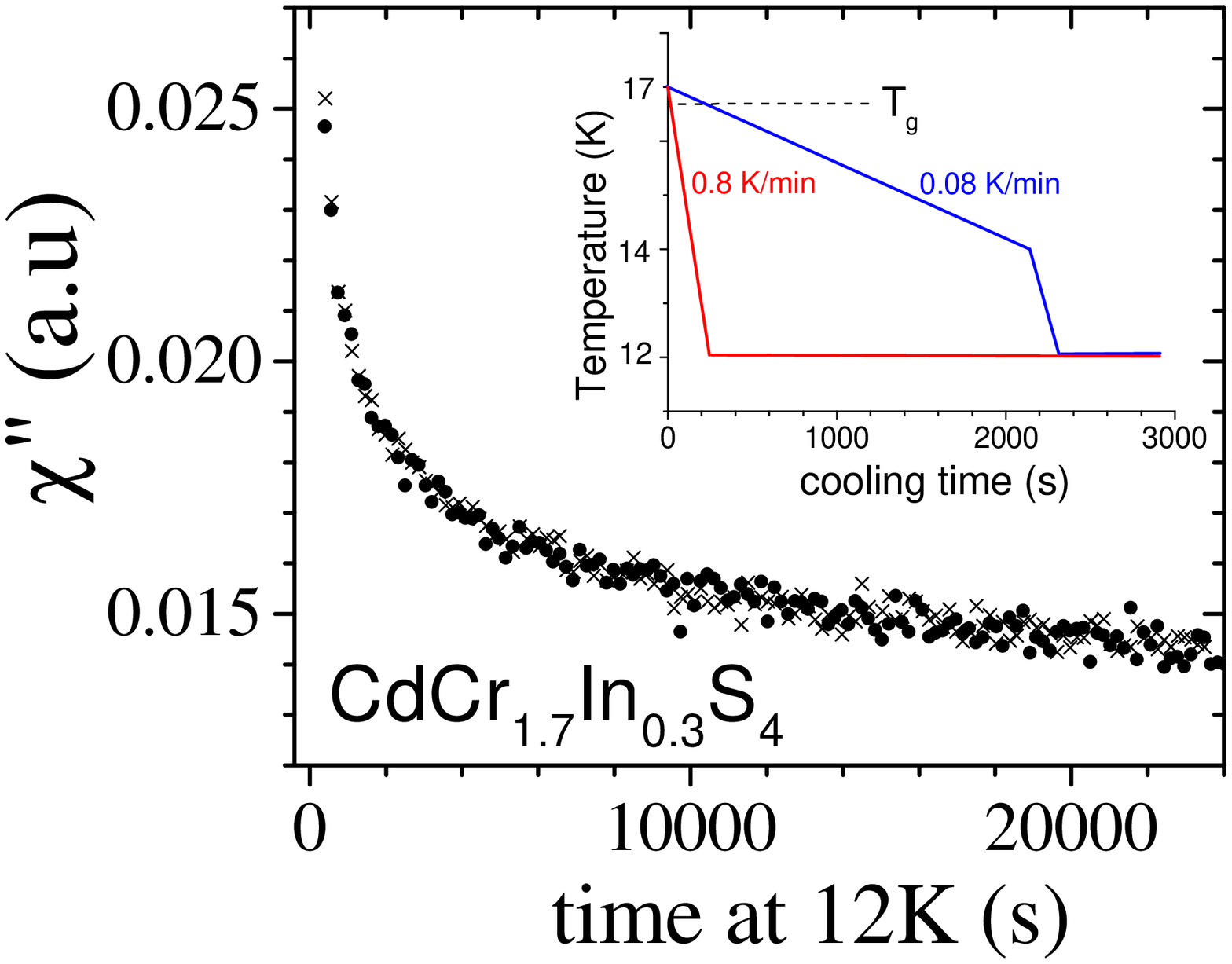,height=7cm}}}
\caption{\small{Effect of the cooling rate (shown in the inset) on $\chi''$ (circles and crosses), showing that 
the value of the cooling rate in the vicinity of $T_g$ is completely
irrelevant. Only the cooling rate during the very last Kelvins actually 
affect the contribution of the aging part of the susceptibility. }} \label{fig1}
\end{figure}

\subsection{Rejuvenation and Memory in temperature cycling experiments}

An even more striking effect has been observed first in spin-glasses \cite{LeFloch,Anders}, and more recently in 
a variety of other systems. Upon cooling, say from $T_1$ to $T_2 < T_1$, the system {\it rejuvenates} and returns to a 
zero age configuration even after a long stop at the higher temperature $T_1$. This is tantamount to saying that the thermal history is irrelevant, as we pointed out
above: stopping at a higher temperature is more or less equivalent to 
cooling the system more slowly. The interesting point, however, is that the
system at the same time remembers perfectly its past history: when
heating back to $T_1$, the value of the a.c. susceptibility is seen to 
match precisely the one it had reached after the first passage at this
temperature, as if the stay at $T_2$ had not affected the system at all: see
Fig. 2. The
paradox is that the system {\it did} significantly evolve at $T_2$, since a 
significant decrease of $\chi''(\omega,t_w)$ is also observed at $T_2$. The
effect would be trivial if no evolution was observed at $T_2$: in this case, one
would say that the system is completely frozen at the lowest temperature, and
then all observables should indeed recover their previous value when the system is heated back. The puzzle comes from the coexistence of {\it perfect memory} on the one hand, and {\it rejuvenation} on the other.\footnote{This rejuvenation 
has often been identified with some kind of `chaotic' evolution of the spin-glass order with temperature. As we shall discuss below, this might be
misleading and we prefer calling this effect `rejuvenation' rather than `chaos'.} Very similar effects have now been seen in different systems: in different spin glasses \cite{memchaos}, some dipolar glasses \cite{Levelut2}, {\sc pmma} \cite{Ciliberto}, and very recently  disordered ferromagnets \cite{Uppsala,Reentrant}. This last case is interesting because the system is 
a so-called reentrant spin-glass: the system is ferromagnetic at high temperatures, and then becomes a spin-glass at lower temperatures. One can
thus compare in detail the aging effects in both phases. The `rejuvenation and memory' turns out to be very similar in both phases, except that 
memory is only partial in the ferromagnetic phase: only when the system is left
at $T_2$ for a rather short time does one keep the memory. In spin glasses, as soon as $T_1 - T_2$ is greater than a few degrees, the memory is kept intact, at
least over experimentally accessible time scales.

\vskip 0.2cm
\begin{figure}
\centerline{\hbox{\epsfig{figure=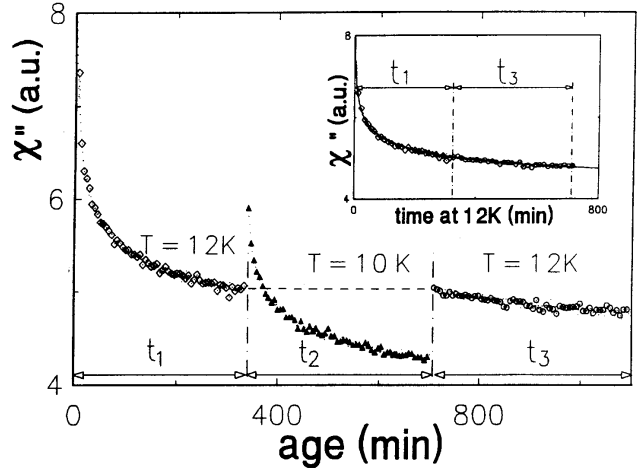,height=7cm}}}
\caption{\small{Effect of a temperature cycle for $T_1=12$ K to $T_2 =10$ K $< T_1$ and
back to $T_1$. One sees that after a long stop at $T_1$, aging is {\it restarted} when the system is cooled to $T_2$, as if it never had stopped at $T_1$ (the central part of the curve would be the same after a direct quench from $T >T_g$ to $T_2$.) At the same time, a perfect {\it memory} of the value of $\chi''$ reached at the end of the stay at $T_1$ is kept somewhere in the system. The inset shows that after ``cutting'' the central part, one recovers 
precisely the usual aging curve at $T_1$. (From \protect\cite{LeFloch}).}} \label{fig2}
\end{figure}

\subsection{Deviations from $t/t_w$ scaling}

We have mentioned above that in many systems such as spin-glasses or polymer
glasses, the aging part of the response function scales approximately as
$t/t_w$, meaning that the effective relaxation time of a system below the
glass transition is set by its {\it age} $t_w$. This scaling is actually 
only approximate. It appears that a better rescaling of the experimental data
is achieved by plotting the data as a function of the difference $(t+t_w)^{1-\mu} - t_w^{1-\mu}$, with $\mu < 1$ \cite{Sitges}. For $\mu=0$, $t_w$ drops out, and this describes a usual time-translation invariant response function, while the limit $\mu \to 1$ 
corresponds to a $t/t_w$ scaling. It is easy to see that the effective relaxation time grows as $t_w^\mu$, i.e. more slowly than the age $t_w$ itself for $\mu < 1$ (`sub-aging'). Note that for dimensional reasons, $t_w^\mu$ should in fact read 
$t_w^\mu \tau_0^{1-\mu}$, where $\tau_0$ is a microscopic time scale. Therefore,
deviations from the simple $t/t_w$ scaling would mean that the microscopic time
scale is still relevant to the aging dynamics, even for asymptotically long
times. Such a behaviour is predicted in some mean-field models of spin-glasses, 
and also in simpler `trap' models, to be discussed below. 

However, one should be very careful in interpreting any empirical value of $\mu$ less than one, because several artefacts can induce such an effective
sub-aging behaviour. A first possible artefact is due to the external field applied to
the system to measure its response. Experimentally, one finds that the higher field (or the external stress), the smaller the value of $\mu$: an external field destroys the aging effect \cite{struik,field}. The extrapolation of $\mu$ to zero field is
somewhat ambiguous, specially because of a second possible artefact, which is
a bad separation between the `fast' relaxation part in Eq. (\ref{Mt}) and the
slow aging part. This comes from the fact that the so-called `fast' part 
$M_\infty(t)$ is not that fast. In spin glasses, it decays as $(\tau_0/t)^\alpha
$, where $\alpha$ is a very small exponent, of the order of $0.05$. Hence, 
this leads to a rather substantial `tail' even in the experimental region
where $t = 10^{12} \tau_0$, which pollutes the aging part and leads to an
effective value of $\mu < 1$. However, even after carefully removing this
fast part, the value of $\mu$ in spin-glasses still appears to be stuck slightly
below the value $1$ \cite{Sitges}. A third reason for this to be so is {\it finite size 
effects}. One expects that for a system of finite size, aging will be
interrupted after a finite time, when the system has fully explored its
phase space. Therefore, after a possibly very long `ergodic' time, the response of
the system has to revert to being time-translation invariant, corresponding to
$\mu=0$. An idea, advocated in \cite{BVH} and recently reconsidered by 
Orbach \cite{Orbach}, is that a sample made of small grains of different sizes
will lead to an effective value of $\mu < 1$ because the ergodic time of some of
the grains enter the experimental time window. 

As a last possible artefact, let us mention the case of dynamical light 
scattering experiments where the signal is recorded {\it while the system
is aging}. In this case, instead of measuring $S_q(t_w+t,t_w)$, one actually
measures:
\be
\tilde S_q(t_w+t,t_w) = \frac{1}{\cal T} \int_0^{\cal T} dt' S_q(t_w+t'+t,t_w+t')
\ee
where $\cal T$ is the integration time needed to obtain a reliable signal. 
It is easy to see that even if $S_q(t_w+t,t_w)$ is a function of $t/t_w$, the
averaging over $t'$ will lead to an effective value of $\mu$ smaller than one,
tending to one only if ${\cal T} \ll t_w$. It is possible that this mechanism 
can explain the value of $\mu \sim 0.5$ determined in \cite{Underwood} for a
colloidal glass, where ${\cal T} \sim t_w$.

Before ending this section, let us finally add an extra comment concerning
the case $\mu > 1$ (`super-aging'). In some simple coarsening models of aging,
discussed below, it can be argued that the relevant scaling variable should
be $\log(t+t_w)/\log(t_w)$ rather than $t/t_w$. This corresponds to an
effective value of $\mu$ greater than one. Other mechanisms leading to $\mu >1$
can be found, see \cite{Takayama} and below.

\section{Simple models of aging}

\subsection{Domain growth in pure systems}

A simple case where aging effects do appear is phase ordering in pure systems \cite{Bray}. Take for example 
an Ising ferromagnet suddenly quenched from high temperatures
to a temperature below the transition temperature. The system then wants to order, and has 
to choose between the up phase and the down phase. Obviously, this takes some time, 
and the dynamics proceeds via {\it domain growth}. After a time $t_w$
after the quench, the typical distance between domain walls is $\xi(t_w)$, which
grows as a power-law of time, i.e. relatively quickly, even for small temperatures. 
A given spin, far from the
domain walls, will thus have to wait for a time $t$ such that $\xi(t_w+t) \sim 2 \xi(t_w)$ 
to flip from -- say -- the up phase to the down phase and decorrelate. For a power-law growth, this means
that the effective relaxation time is of the order of $t_w$ itself, corresponding to $\mu=1$.
This behaviour is confirmed by several exactly soluble models of
coarsening, such as the Ising model on a chain, or the `spherical' model, where
one can compute explicitly the correlation function to find \cite{Bray}:
\be
C(t_w+t,t_w) = C_\infty(t)+C_{\sc ag}(\frac{t}{t_w}).
\ee
One can also compute several other quantities, such as the effect of an external magnetic field $h$ on the aging properties. One then sees that when $\xi(t_w)^{-1}$ is smaller than $h$,
the driving force due to the curvature of the domain walls is superseded by the driving force due
to the external field \cite{CuDean}. In this situation, the favoured phase quickly invades the whole system and
aging is stopped. This leads, for very small fields and moderate time scales, to an effective value
of the exponent $\mu < 1$, as discussed in section 2.4. The main quantity of interest to compare
with experiments, however, is the response function. The result is that the aging part of the 
response function vanishes as $\xi(t_w)^{-1}$ as $t_w$ becomes large. In terms of the a.c.
susceptibility, one finds that \cite{CuDean,Jorge}:
\be
\chi''(\omega,t_w) = \chi''_\infty(\omega) + \xi(t_w)^{-1} \chi''_{\sc ag}(\omega t_w).
\ee
Intuitively, this result means that the aging part of the susceptibility only comes from the 
domain walls, while the spins in the bulk of domains contribute to the stationary part 
$\chi''_\infty(\omega)$. Since the density of spins belonging to domain walls decreases as 
$\xi^{d-1}/\xi^d = \xi^{-1}$, the aging contribution decreases with time as the density of walls.

Domain growth in pure systems is driven by surface tension and does not require thermal
activation; the aging effects in these systems are therefore hard to detect experimentally, since the
typical size of the domains soon reaches its maximum value [set either by the size of the system
of by magneto-static (or other) considerations.] Similarly, one does not expect the cooling rate to
have a major influence on the coarsening of the system.

\subsection{Domain growth in random systems}

\vskip 0.2cm
\begin{figure}
\centerline{\hbox{\epsfig{figure=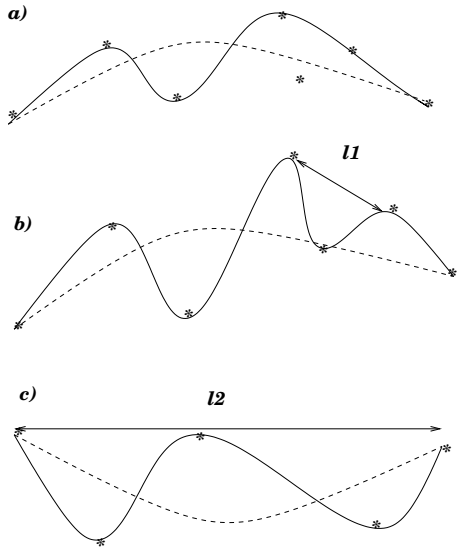,height=6cm,angle=270}}}
\caption{\small{Schematic evolution of a pinned object on well separated time scales. On scale
$t(\ell_1)$, the object reconforms by flipping a small portion of size $\ell_1$ from one
favourable configuration to another (a $\to$ b). On a much longer time scale $t(\ell_2) \gg t(\ell_1)$, the conformation on scale $\ell_2$ (dotted lines) has evolved (b $\to$ c). The dynamics of the short wavelengths happens
on a time scale such that long wavelengths are effectively frozen.}} \label{fig3}
\end{figure}

More interesting is the situation in disordered ferromagnets (for example in the presence of quenched
random fields or random bonds). In this case, the impurities act as pinning sites for the domain walls.
The problem of elastic objects (such as domain walls, but also vortices in superconductors, dislocations,
etc.) pinned by random impurities has been the subject of intense work in the past decade, both from a
static point of view (where the typical equilibrium conformation of such objects is investigated) or
to understand their dynamics (relaxation to equilibrium, response to an external driving force, creep 
and depinning transition) -- for reviews, see \cite{Fisher,LeDou,Nattermann,Yoshino}. Actually, these systems constitute `baby' spin glasses: frustration is
present because of the competition between pinning, which tends to distort the structure, and elasticity
which tends to reduce the deformations. The main result of the theory is
the appearance of a typical pinning energy $E_{p}(\ell)$ associated to the linear size $\ell$ of the piece
of domain wall that attempts to reconform. This energy scale {\it grows as a power} of $\ell$:
\be
E_{p}(\ell) \sim \Upsilon(T) \ell^\theta,
\ee
where $\Upsilon$ is a (temperature dependent) energy scale, and $\theta$ an exponent which depends
on the dimensionality of the structure (1D for dislocations, 2D for domain walls, etc.) and on
the correlations of the pinning field. Using a very naive Arrhenius law for thermal activation, this
means that the typical time associated to reconformation events that occur on a scale $\ell$ take
a time: 
\be\label{tell}
t(\ell) \sim \tau_0 \exp\left[\frac{\Upsilon(T) \ell^\theta}{T}\right].
\ee
Equivalently, the size over which the system can equilibrate after a time $t_w$ only grows {\it logarithmically}:
\be\label{growth}
\ell(t_w) \sim \left[\frac{T}{\Upsilon(T)} \log(\frac{t_w}{\tau_0})\right]^{1/\theta}
\ee
In particular, the typical size of the growing domains in expected to grow logarithmically with time, 
that is, {\it extremely slowly}.\footnote{Actually, for domain growth, the exponent of the logarithm 
is probably larger than $1/\theta$, but this does not matter for the present qualitative discussion. In particular, one expects the aging part of the correlation function to 
be a function of $\log(t_w+t)/\log(t_w)$.}
This very slow growth means that the density of domain walls only decays slowly with time, and therefore
that the aging contribution to the susceptibility is still significant even after macroscopic times. 

Although very simple, the exponential relation between length scales and time scales given by
Eq. (\ref{tell}) has far-reaching consequences: the dynamics becomes, in a loose sense, hierarchical.
This is illustrated in Fig. \ 3. The object evolves between metastable configurations which differ 
by flips of regions of size $\ell_1$ on a time $t(\ell_1)$ that, because of the exponential 
dependence in Eq. (\ref{tell}), is much shorter than the time needed to flip a region of size $\ell_2 > \ell_1$.
Therefore, the dynamics of the short wavelengths happens on a time scale such that long wavelengths are effectively frozen. As we shall explain below, this feature is, in our eyes, a major ingredient to understand the coexistence of rejuvenation and memory. Another important consequence is the fact that domain growth becomes a very intermittent process: once an event on the scale of the domain size $\xi$ has happened,
the details of the conformation on scales $\ell < \xi$ start evolving between nearby metastable states, 
while the overall pattern formed by the domains on scale $\xi$ hardly change.
 
The equation (\ref{tell}) also allows one to define a very important quantity which we call,
by analogy with the glass temperature $T_g$, the `glass length'
$\ell_g$, through $\Upsilon(T) \ell_g^{\theta} = {\cal A} T$, introduced in this context in \cite{BD,BBM,Yoshino}. The factor ${\cal A}$ is rather arbitrary; the choice ${\cal A}=35$ corresponds to a time of $1000$ seconds if $\tau_0=10^{-12}$ seconds. In analogy with the glass temperature $T_g$, one sees that length scales larger than $\ell_g$ cannot be equilibrated on reasonable time scales, while length scales
smaller than $\ell_g$ are fully equilibrated.  Qualitatively speaking, the equilibrated modes contribute to the stationary part of the correlation and/or response function, while the glassy modes $\ell > \ell_g$ contribute
to the aging part. Therefore, the strong hierarchy of time scales induced by the exponential activation 
law allows equilibrated modes and aging modes to coexist. 

Finally, it is easy to understand that the logarithmic growth law, Eq. (\ref{growth}), leads to a strong {\it cooling rate} dependence of the typical size of the domains \cite{Levelut}:
since the growth law is essentially that of pure systems as long as $\xi \ll \ell_g(T)$, a longer time spent at higher temperatures (where $\ell_g$ is large) obviously allows the domains to grow larger before getting pinned at lower temperatures.

\subsection{Diffusion in a random potential 1: The Sinai model} 

It is useful to consider a toy model for the dynamics of a pinned domain wall by considering the motion of
a {\it point particle} in a random potential. This can be thought of as a reduction of the problem to the
dynamics of the center of mass of the pinned object. From the study of these pinned objects, it is known 
that reconformations on scale $\ell$ typically change the position of the center of mass $X$ by an amount
$\propto \ell^\zeta$, where $\zeta$ is a certain exponent, analoguous to the exponent $\theta$ defined above.
Since the energy changes by an amount $\Upsilon \ell^\theta$, the statistics of the random potential $V(X)$
acting on the center of mass $X$ must be such that:
\be
\langle [V(X)-V(X')]^2 \rangle \propto \Upsilon^2 |X-X'|^{2\theta/\zeta} \qquad |X-X'| \ll L^\zeta,
\ee
where $L$ is the total transverse size of the object. For larger distances, $\langle [V(X)-V(X')]^2 \rangle$
saturates to a finite value, since the impurities encountered by the pinned object become uncorrelated.

An interesting example is provided by a {\it line} in a plane (i.e. a domain wall in two dimensions) in the
presence of impurities. In this case, the exponents $\zeta$ and $\theta$ are exactly known and read:
$\zeta=2/3$, $\theta=1/3$, leading to $\langle [V(X)-V(X')]^2 \rangle \propto \Upsilon^2 |X-X'|$. If one 
assumes that the statistics of $V(X)$ is Gaussian, then the potential $V(X)$ is a random walk, and the model
under consideration is precisely the well known Sinai model, for which a large number of analytical results are known (for reviews, see \cite{Annals,BG,FML}).

One can also define in this model a `glass scale' $X_g$ such that, at temperature $T_1$, 
that:
\be
\Upsilon(T_1) \sqrt{X_g} = {\cal A} T_1 \qquad \mbox{or} \qquad 
X_g=\left(\frac{{\cal A}T_1}{\Upsilon(T_1)}\right)^2,
\ee
where we have taken into account a possible temperature dependence of $\Upsilon$.
For $X \ll X_g$, the diffusion motion is quasi-free, and the dynamics is 
`fast' (diffusive). For times
corresponding to distances larger than $X_g$, the energy barriers strongly 
impede the motion, and lead to a slow logarithmic sub-diffusion:
\be
t \sim \tau_0 \exp[\frac{V(X)}{T_1}] \longrightarrow X(t) \sim X_g \log^2 (\frac{t}{\tau_0}).
\ee
More precisely, all the particles initially lauched at $X=0$ will, after time
$t$, be located in the deepest energy well available at time $t$, which is at a distance $\sim \log^2 t$ from the initial point. The relative distance between 
these particles, however, does not grow with time. The deepest trap available
is so much more favourable than the others that the relative distance between all particles remains typically of order $X_g$: this is the Golosov phenomenon
(see \cite{Gol,BG,Compte}). Within the deepest well, on the other hand, the probability is roughly uniform, since the energy landcape is shallow in 
comparison with $T_1$. 
One can actually argue, using the beautiful results of \cite{FML}, that the (intra-well) response of the 
particle to a small oscillating external field should behave as:
\be
\chi_{\sc ag}(\omega,t_w) \sim \frac{\log^3 t_w}{\omega t_w}
\ee
where $t_w$ is the time since the quench from very high temperatures. 
This 
shows that within
this simplified model, the response is not exactly a function of $\omega 
t_w$, but that there are logarithmic corrections. This means that the scaling 
variable is again, over a limited range of $t_w$, of the form $\omega 
t_w^\mu$ with $\mu < 1$. In Fig. \ 4, we show the results of a numerical 
simulation performed with H. Yoshino \cite{YB}, where $\chi(\omega,t_w)$ is 
computed for the Sinai model at various frequencies, as a function 
of $t_w$. We have shown the slope $1/t_w$ for comparison, and rescaled the 
different curves
using the value $\mu=0.9$ in the inset. 

\vskip 0.2cm
\begin{figure}
\centerline{\hbox{\epsfig{figure=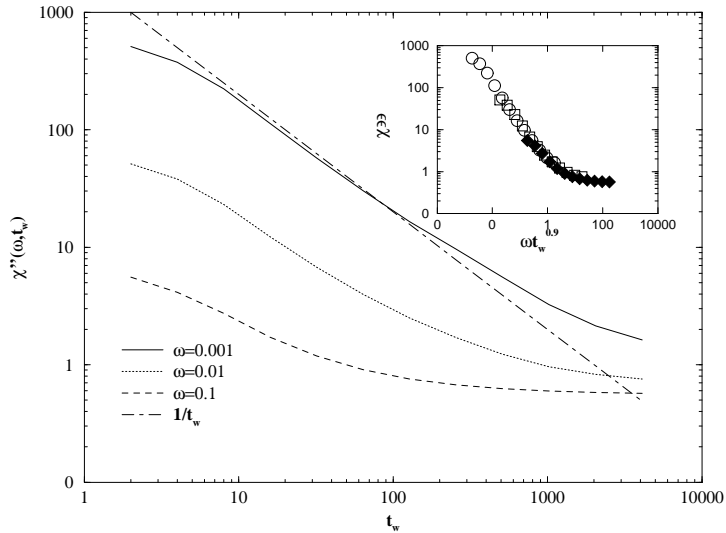,height=6cm,angle=270}}}
\caption{\small{Dissipative part of the response function to a small 
oscillating field in the Sinai model, as a function of waiting time, for different frequencies. The slope $1/t_w$ is shown for comparison. In the inset, we show that these curves can be rescaled using an
effective value of $\mu=0.9$.}} \label{fig4}
\end{figure}

Now, the very interesting property of the Sinai model is the fact that the potential $V(X)$
is strictly self-affine. This means that the statistics of the potential at 
small scales
is identical, up to a scaling factor, to the statistics of the potential at 
larger scales (see Fig. \ 5). Therefore, when the temperature is lowered from
$T_1$ to $T_2 < T_1$ after a time $t_{w1}$, the particle has to restart its search for the most 
favourable well, much as it had done when the temperature first reached $T_1$
from high temperatures. Since the probability distribution within the well is uniform at the
moment of the second temperature change, it indeed corresponds, effectively, to a high temperature quench. The point however is that the particles cannot leave the deep well
they had reached at $T_1$ before a time exceedingly long compared to $t_{w1}$.
This time $t_{w2}^*$ is the time needed to overcome the depth of the well 
reached at $t_{w1}$, and is given by:
\be
t_{w2}^* = \tau_0 (\frac{t_{w1}}{\tau_0})^\beta \qquad \mbox{with} \qquad \beta=\frac{T_1 \Upsilon(T_2)}{T_2 \Upsilon(T_1)}.
\ee
Consider the case where the pinning energy vanishes above a certain temperature
$T_c$, as $\Upsilon(T) \sim (T_c-T)^\omega$, with a certain exponent $\omega$. (This is the case, for example, for domain walls near the Curie temperature).
Then taking $\omega=1$, $T_1=0.9 T_c$ and $T_2 = 0.8 T_c$, one finds that 
$\beta = 2.25$. Therefore, if $\tau_0 = 10^{-12}$ second, and $t_{w1}=1$ second,
one finds that $t_{w2}^*$ is astronomically long, equal to $10^{15}$ seconds!
In other words, the particle is completely trapped, even when the temperature only
changes by $10\%$. 

\vskip 0.2cm
\begin{figure}
\centerline{\hbox{\epsfig{figure=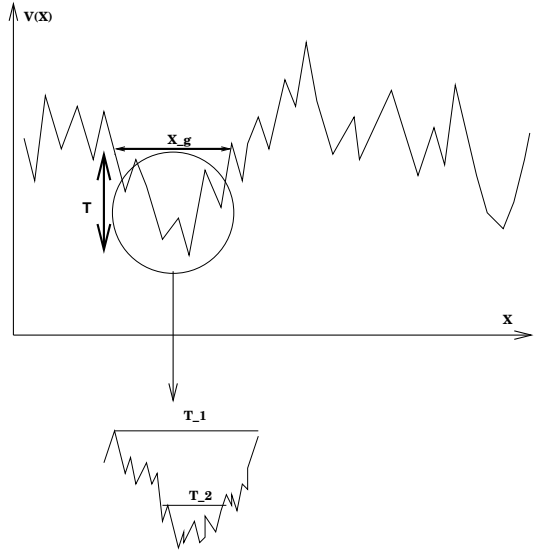,height=7cm,angle=270}}}
\caption{\small{The potential $V(X)$ in the Sinai model is self-affine. A
zoom of a small portion of it, of size $X_g(T_1)$, reveals the same
statistical features at a smaller scale. Correspondingly, the
dynamics at a smaller temperature $T_2$ will exhibit 
properties after a quench from $T_1$ very similar to the initial dynamics after
a quench from high temperature. On the other hand, the dynamics at
$T_2$ mostly takes place within the valley reached at $T_1$.}} \label{fig5}
\end{figure}

Therefore, this model provided a tantalizing scenario for the `rejuvenation
and memory' effect: as the temperature is cooled down, new details of the 
potential appear, in a self-similar manner, and the aging dynamics over the
barriers starts afresh. On the other hand, the particle remains effectively 
forever in the well that it had reached at $T_1$. Therefore, when the
system is heated back to $T_1$, perfect memory is recovered. This scenario
is very similar to the one advocated for spin glasses on the basis of a
hierarchical landscape, inspired by Parisi's mean field solution \cite{Vincent,BD}. 
We believe that the Sinai model 
offers a precise basis for such a picture. Note that there is no `chaos'
involved in this model: rejuvenation occurs because previously equilibrated
modes are thrown out of equilibrium, but in the very same energy landscape.

\subsection{Diffusion in a random potential 2: The trap model} 

The above model assumes that the random potential is Gaussian, with a correlation 
function compatible with direct scaling arguments, which lead to $V(X) \sim X^{\theta/\zeta}$. A slightly different picture emerges from the replica variational
theory which predicts a highly {\it non-Gaussian} effective pinning potential acting on
the center of mass of the pinned object. As detailed in \cite{BBM,Gumbel}, the potential is
a succession of local parabolas (corresponding to locally favourable configurations, or `traps') matching at singular points (see also \cite{Feigelmann}). The full `replica symmetry breaking' scheme needed to reproduce
the correct scaling of the potential (i.e. $V(X) \sim X^{\theta/\zeta}$) means that the
potential is actually a hierarchy of parabolas within parabolas, etc. 
This hierarchy directly corresponds to the existence of different length scales $\ell$, each one characterized by an energy scale $E_\ell \sim \Upsilon \ell^\theta$. An important difference
with the above Sinai model is the statistics of the depth of the different valleys of the
pinning potential: the prediction of the replica approach is that the deepest valleys obey 
the so-called Gumbel extreme value statistics:
\be
P_\ell(E) \sim_{E \gg E_\ell} \frac{1}{E_\ell} \exp(-\frac{E}{E_\ell}).
\ee
Assuming that the time needed to hop out of a trap of depth $E$ is $\tau=\tau_0 \exp(E/T)$,
one finds that the above exponential distribution of trap depths induces a {\it power-law}
distribution of trapping times:
\be\label{traps}
P_\ell(\tau) \sim_{\tau \gg \tau_0} \frac{\tau_0^{x_\ell}}{\tau^{1+x_\ell}} \qquad x_\ell=\frac{T}{E_\ell}\sim (\frac{\ell_g}{\ell})^\theta.
\ee
The model of a random walk between independent traps such that their release time is 
given by Eq. (\ref{traps}) has been investigated in details in \cite{MB,BD}, and more recently, in the context of the rheology of soft glassy materials, in \cite{Sollich}.
The interesting result is that the dynamics of this simple model is time-translation
invariant as long as $x_\ell > 1$ (high temperature phase, corresponding to small
length scales), but becomes {\it aging} when the
average trapping time diverges, i.e. when $x_\ell < 1$ (large length scales). In this case, the dynamics becomes extremely intermittent, since the system spends most of its time in
the deepest available trap.

One can compute, within this simple model, the a.c. susceptibility (or frequency dependent elastic modulus) to find
\cite{MB,BD,Sollich}:\footnote{Note that the result quoted above for the Sinai model
formally corresponds to $x_\ell=0$, a result that could have been anticipated \cite{Annals}, up to logarithmic corrections.}
\be
\chi_\ell''(\omega,t_w) = A(x_\ell) (\omega\tau_0)^{x_\ell-1};\ (x_\ell > 1) \quad \mbox{and}\quad
\chi_\ell''(\omega,t_w) = A(x_\ell) (\omega t_w)^{x_\ell-1};\ (x_\ell < 1).
\ee
The case where all length scales evolve in parallel therefore leads to a total susceptibility given by $\chi''(\omega,t_w)=\chi''_\infty(\omega)+\chi''_{\sc ag}(\omega t_w)$ with:
\be\label{chitrap}
\chi''_\infty(\omega)= \sum_{\ell < \ell_g} {\cal G}(\ell)A(x_\ell) (\omega\tau_0)^{x_\ell-1} \qquad  
\chi''_{\sc ag}(\omega t_w)= \sum_{\ell > \ell_g} {\cal G}(\ell)A(x_\ell) (\omega t_w)^{x_\ell-1},
\ee
where $\cal G$ is a `form factor' counting the number of available modes at scale $\ell$.
A very interesting consequence of Eq. (\ref{chitrap}) is that in the low frequency,
long waiting time limit (more precisely when $\omega \tau_0 \ll 1$ and $\omega t_w \gg 1$),
the sum over $\ell$ is dominated by the region $\ell \sim \ell_g$, for which $x_\ell \sim 1$. For a fairly general function $\cal G$, one expects both the stationary and aging part of $\chi''$ to behave asymptotically as $1/\log \omega$. This can be translated, using the fluctuation
dissipation theorem, into a noise spectrum $S(\omega) \propto 1/\omega \log \omega$, i.e.
a so-called $1/f$ noise, ubiquitous in glassy systems. This mechanism suggests that
$1/f$ noise should generically exhibit a slowly evolving component. 
Following the same argument, one also
expects that the aging contribution to $\chi''$ decays as a small power of $t_w$, eventually
reaching a $1/\log(\omega t_w)$ behaviour for $\log(\omega t_w) \gg 1$. As mentioned above (see Eq. (\ref{sgaging})), data on spin glasses
typically give $1 - x_\ell \sim 0.1 - 0.3$ for $\omega t_w$ in the range $1-10000$.

The strictly hierarchical nature of the landscape also leads, by construction, to the `rejuvenation and memory' effect \cite{Vincent,BD}: when the temperature is reduced from $T_1$ to $T_2$, the `glass length scale' moves down from $\ell_{g1}$ to $\ell_{g2}$. Modes corresponding to
$\ell_{g2} < \ell < \ell_{g1}$, which were equilibrated at $T_1$, become aging at $T_2$ (hence the `rejuvenation'), while the modes
such that $\ell > \ell_{g1}$, which were aging at $T_1$, become effectively frozen at $T_2$
(hence the `memory'). This is illustrated in Fig. \ 6. 

\vskip 0.2cm
\begin{figure}
\centerline{\hbox{\epsfig{figure=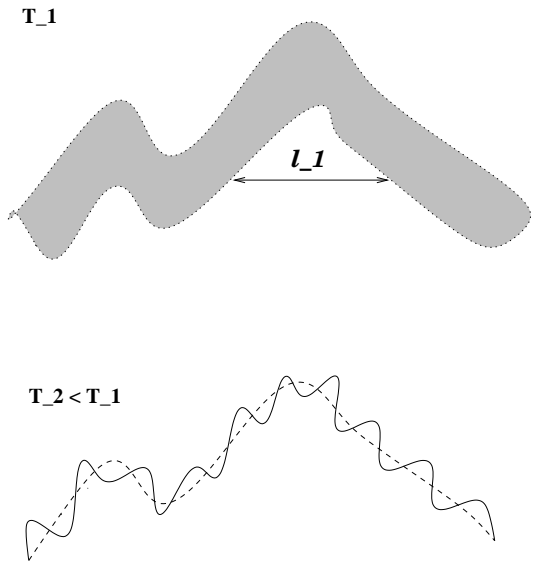,height=7cm,angle=270}}}
\caption{\small{Scenario for rejuvenation and memory for a pinned object:
at temperature $T_1$, the length scale $\ell_1$ is slowly evolving, while the 
smaller length scales are `fast', and thus lead to a `blurred shaped' object
if one looks at it on the time scale $t(\ell_1)$. Upon cooling down to
$T_2$, the smaller length scales start `condensing': the object is pinned on smaller length scales.}} \label{fig7}
\end{figure}

Let us finally discuss how values of $\mu \neq 1$ can arise within the trap 
model. If one assumes that the traps visited during the evolution of the 
system are all different, as 
implicitly done above, then $\mu=1$. This is not true if, for example, the 
geometry of the trap connectivity is of low dimension, for example if the 
traps are on a one-dimensional structure. In this case, one can show 
\cite{Maass} that $\mu=1/(1+x)$ for $x < 1$. One can also
look at simpler models, when a particle makes directed hops on a line where 
traps have a deterministic lifetime $\tau(n)$ which grows with the distance 
$n$ to the origin. The position $N(t_w)$ of the particle at time $t_w$ is 
therefore given by:
\be
t_w = \sum_{n=1}^N \tau(n),
\ee
since we have assumed that the walk is directed towards $n >0$. The 
subsequent evolution of the particle, for a time $t \ll \tau(N)$, will be a 
function of $t/\tau(N)$. Taking $\tau(n)
\propto n^\beta$ immediately leads to $\mu=\beta/(\beta+1) < 1$. If $\tau(n)$ grows faster 
than exponentially with $n$, then of finds that the effective value of $\mu$ 
is larger than one, 
because of the presence of logarithmic corrections which lead to superaging.

\section{Back to experiments}

Equipped with the above theoretical ideas, one can return to the experimental
results presented in the first sections, and see how far one can go in their
interpretation.
\subsection{Disordered Ferromagnets}

From the discussion above, one expects that aging in disordered ferromagnets 
can be
understood in terms of the superposition of slow domain growth, and domain 
walls 
reconformations in the pinning field created by the disorder. 
Correspondingly, the aging part of the susceptibility is expected to behave 
as:
\be
\chi''_{\sc ag}(\omega,t_w)=\frac{1}{\xi(t_w)} 
\left[\chi''_{w\infty}(\omega)+ \chi''_{w\sc{ag}}(\omega,t_w)\right],
\ee
where $\xi(t_w)$ is the (slowly growing) domain size, $\chi''_{w\infty}$ the 
stationary
mobility of the domain walls at frequency $\omega$, and $\chi''_{w\sc{ag}}$ the aging contribution 
coming from
the reconformation modes, expected to scale as $\omega t_w$. This expression 
accounts well for the observations made in 
random ferromagnet like systems, such as the ones studied in 
\cite{Levelut,Reentrant}, in particular:
\begin{itemize}
\item
The aging part of the response is quite sensitive to the cooling rate, and 
decreases when 
cooling is slower and/or when the waiting time increases. This dependence is 
logarithmic, 
and directly reflects the behaviour of $\xi(t_w)$.
\item For small frequencies, $\chi''_{w\sc{ag}}$ dominates and one observes 
an approximate 
$\omega t_w$ scaling of the aging part of $\chi$, up to log-corrections coming 
from the $1/\xi$ factor. On the other hand, for 
larger frequencies, the reconformation contribution becomes negligible and 
the $\omega t_w$ scaling breaks 
down completely, as observed in \cite{Levelut}.
\item 
One observes rejuvenation and memory, induced by the slow reconformation of 
the walls. However, memory is only recovered if the time spent at the lower 
temperature $T_2$ is
short enough, such that the overall position of the domain walls has not 
changed significantly. In the other limit, i.e. when the walls can move 
substantially, the impurities interacting with the walls are completely 
renewed, and memory is lost \cite{Reentrant}.

The fact that aging in glycerol bears some resemblance with aging in disordered
ferromagnets suggests that some kind of domain growth might also be present
in structural glasses. The precise nature of this domain growth is however
at this stage not very clear, but is certainly a very interesting subject to explore further.

\end{itemize}
\subsection{Spin-glasses}
The interpretation of the aging experiments in spin-glasses is not as 
transparent; this is directly related to the fact that
the correct theoretical picture for spin-glasses in physical dimensions (as opposed
to mean-field) is 
still very much controversial. The simplest description is the droplet 
theory, where one essentially assumes
that a spin-glass is some kind of `disguised ferromagnet', in the sense that 
there are only
two stable phases for the system, that one can (by convention) call `up' and 
`down'. The dynamics of the system after a quench can then be again thought 
of in terms of domain growth in a disordered system, with the difference that 
the `pattern' that is progressively
invading the system is itself random. Correspondingly, the energy of a `domain' grows with its size $\ell$ as $\ell^\theta$, where $\theta$ is smaller than the value $d-1$ which holds in a pure ferromagnet. 
This is the scenario proposed (in a dynamical context) by Fisher 
and Huse \cite{FH}, and further investigated by Koper and Hilhorst \cite{KH}. 
However, this scenario immediately stumbles on a first difficulty, in that it 
would predict
a very strong cooling rate dependence of the susceptibility which, as 
shown in Fig. \ 1, is definitely not observed.

A way out of this contradiction is to argue that even if {\it at any given
temperature}, there are only two stable phases in competition (i.e. one pattern
and its spin reversed), the favoured pattern changes chaotically with temperature. More precisely, if the temperature changes by $\Delta T$, then the
precise relative arrangement of the spins is preserved for lengths scales less
than a `chaos' length $\ell_{\Delta T} \propto \Delta T^{-y}$, and completely
destroyed at larger length scales \cite{Chaos}. (Here, $y$ is a new exponent related to 
$\theta$.) If this is the case, then all the aging achieved at higher temperatures is useless to bring the system closer to equilibrium at $T_1$, where domain growth has to restart from scratch. This also explains how the
system rejuvenates upon a small temperature change. 

The problem with this interpretation is that: (i) chaos with temperature has not been convincingly established neither theoretically nor numerically. In particular, there seems to be no `chaos' with temperature in the mean-field Sherrington-Kirkpatrick model \cite{Billoire}; and (ii) that if the evolution at $T_2$ consists
in growing new domains of the $T_2$ phase everywhere in space, it is difficult to imagine how this does not destroy completely the correlations built at 
temperature $T_1$. A way to do this would be to say that the new $T_2$-phase  
only nucleates around particular nucleation sites and grows very slowly, in 
such a way that a substantial region of space is still filled by the $T_1$ 
phase. However, this would mean that it is impossible to describe the dynamics
of the system in terms of two phases only, as assumed in the droplet model. At any temperature, the configuration would necessarily
be a mixture of all the phases corresponding to the previous temperatures 
encountered during the thermal history of the system \cite{memchaos}.

The `droplet' theory is radically different from the picture emerging from the
Parisi solution of the mean-field Sherrington-Kirkpatrick model \cite{MPV}. There, one
can show that the number of `phases' in which the system can organize is very large. More precisely, there are configurations of nearly equal energy which differ by the flip of a finite fraction of the total number of spins. A consistent picture for how this scenario applies for finite dimensional 
spin-glasses is however still missing. A recent interesting suggestion made by Houdayer and Martin is that these different phases differ by the reversal of
large non-compact, sponge like objects \cite{HM}. These objects have a linear dimension 
equal to the size of the system, but are not space-filling. Rather, their 
boundary separates an `interior' from an `exterior' which form a bi-continuous
structure. If this is the case, then by definition this boundary is a kind
of domain wall with zero surface tension. This is crucial in the sense that 
these `domain walls' can hop from one metastable configuration to another
(much as in a disordered ferromagnet) but with {\it no overall tendency to 
coarsen}.

In other words, the ordered phase of spin-glasses is, in a sense, full of 
{\it permanent} domain wall-like 
defects. \footnote{These permanent defects are probably very similar to the  `active' droplets of Fisher and Huse.} These domain walls can only be precisely defined in reference to the 
true ground 
state of the system; their physical reality should however be thought of as 
particularly `mobile' regions of spins; the precise position of these `walls' 
define the possible metastable
states of the sample. 

After quenching the system to low temperatures, the system coarsens to get 
rid of the excess
intensive energy, as has been beautifully demonstrated numerically in 
\cite{Takay2} (see also \cite{Rome,Joh}). However, the state left behind is still 
a `soup' of walls with zero surface tension. The density of these walls is 
large and does not decay with time, and therefore  provide the major 
contribution to the aging signal. The rejuvenation and memory effect can be 
understood in terms of the progressive quenching of smaller and smaller 
length scales of those walls as the 
temperature decreases, as we argued above for disordered ferromagnets. This 
interpretation is actually motivated by the fact that the temperature cycling 
experiments in the ferromagentic phase and in the spin-glass phase of a 
single `reentrant' spin-glass sample reveal very
similar features \cite{Reentrant}. The only difference is that memory is perfect in 
spin-glasses, {\it as if}
no coarsening was present.

Let us insist once more on the fact that the above mechanism for rejuvenation 
is very different
from the `chaos' hypothesis. Within the trap model, aging is the phenomenon 
which occurs when the Boltzmann weight,
initially uniformly scattered among many microstates, has to `condense' into a 
small fraction
of them. It is the dynamic counterpart of the entropy crisis transition that 
takes place in the
Random Energy Model \cite{REM,Gumbel}.

Obviously, the above discussion is very speculative and a deeper 
understanding of aging in
spin-glasses is very much needed. We however believe that the idea that the 
ordered phase of a spin-glass contains a large number of pinned, zero tension 
walls, which reconform in their disordered landscape, is a useful picture.

\section{Conclusion}

In these lectures, we have tried to review the most striking experimental results
on aging in a variety of disordered systems, which reveal similar features but
also important differences. We have argued that a generic model that reproduce
many of these features is that of {\it pinned defects} in a disordered environment. 
The fact that energy barriers grow with the size of the reconforming regions 
immediately leads to a strong hierarchy of time scales. In particular, long wavelength aging modes and short wavelength equilibrated modes coexist in the
system and offer a simple mechanism to explain the rejuvenation/memory effect. 
These properties can be discussed within simplified models where the dynamics whole pinned object is reduced to that of its center of mass in a disordered potential. 
The main difference between random ferromagnets and spin-glasses seems to lie in the
fact that while domains slowly grow in the former case (and therefore progressively
reduces the density of domain walls), the fraction of `domain walls' in spin-glasses
appears to remain constant in time. 

We have not attempted to discuss here the dynamical mean-field models, which have been
much studied recently (for a review, see \cite{Review}). These models are also 
able to reproduce many of the interesting features of aging, including the rejuvenation/memory effect \cite{CuKLast}. Furthermore, these models
allow one to make precise predictions on the possible violations of the Fluctuation-Dissipation Theorem \cite{CuK,CuKP}, and the relation between this violation and the 
non trivial overlap distribution function which appears in the static solution of these models \cite{Pq}. Finally, the exact dynamical equations of these models at high temperature are very similar to those of the Mode-Coupling Theory for fragile glasses. Therefore, one can study by analogy the extension of the Mode-Coupling equations to the glass phase,
and obtain, within this framework, interesting results on the aging properties of 
fragile glasses \cite{MCA}. However, the relevance of these mean-field theories to finite
dimensional systems is not obvious, especially at low temperatures. This is because
these models actually describe the diffusion of a single particle in a very high
dimensional disordered potential \cite{KL,YIKS}. In high dimensions, the particle is never trapped in the bottom of a valley: there are always directions to escape.\footnote{This is true at least for discontinuous spin-glasses. The geometrical interpretation for `continuous' spin-glasses is
less clear, though. \cite{Review}} The aging dynamics is dominated by the fact that the average number of unstable directions 
is decreasing with time, not by the fact that typical energy barriers are growing with time \cite{Cavagna}. The inclusion of true activated effects in these mean-field (or Mode-Coupling)
equations is still very much an open problem.

\section*{Acknowledgements} The ideas presented here owe a lot to many collegues and friends with whom I have discussed these matters over the years, in particular F. Alberici, L. Cugliandolo, D. Dean, V. Dupuis, P. Doussineau, J. Hammann, J. Kurchan, A. Levelut, M. M\'ezard, P. Nordblad, E. Vincent and H. Yoshino. Interesting discussions with J. L. Barrat, A. Bray, M. Cates, D. S. Fisher, J. Houdayer, W. Kob, Ph. Maass, E. Marinari, O. Martin, R. Orbach, E. Pitard and P. Sollich must also be acknowledged. Finally, I want to thank Mike Cates for inviting me to give these lectures in St Andrews, and for a most enjoyable and fruitful collaboration over the years.

\end{document}